\documentclass[runningheads]{llncs}
\usepackage[T1]{fontenc}
\usepackage{graphicx}
\usepackage{multirow}
\usepackage{amsmath}
\usepackage{hyperref}

\usepackage[misc]{ifsym}

\begin{document}

\title{A Situation-aware Enhancer for Personalized Recommendation}
\author{Jiayu Li\inst{1}\orcidID{0000-0002-6351-897X} 
\and Peijie Sun\inst{1}\orcidID{0000-0001-9733-0521} 
\and Chumeng Jiang\inst{1}\orcidID{0009-0005-6247-2497} 
\and Weizhi Ma\inst{2}\Letter\orcidID{0000-0001-5604-7527}
\and Qingyao Ai\inst{1}\orcidID{0000-0002-5030-709X}
\and Min Zhang\inst{1}\Letter\orcidID{0000-0003-3158-1920}}
\authorrunning{J. Li et al.}
\institute{Quan Cheng Lab; DCST, Tsinghua University 
\email{z-m@tsinghua.edu.cn}
\and
Institute for AI Industry Research, Tsinghua University
\email{mawz@tsinghua.edu.cn}
}

\maketitle  

\begin{abstract}
When users interact with Recommender Systems~(RecSys), current situations, such as time, location, and environment, significantly influence their preferences. Situations serve as the background for interactions, where relationships between users and items evolve with situation changes. However, existing RecSys treat situations, users, and items on the same level. They can only model the relations between situations and users/items respectively, rather than the dynamic impact of situations on user-item associations~(i.e., user preferences). In this paper, we provide a new perspective that takes situations as the preconditions for users' interactions. This perspective allows us to separate situations from user/item representations, and capture situations' influences over the user-item relationship, offering a more comprehensive understanding of situations. Based on it, we propose a novel Situation-Aware Recommender Enhancer~(SARE), a pluggable module to integrate situations into various existing RecSys. Since users' perception of situations and situations' impact on preferences are both personalized, SARE includes a Personalized Situation Fusion~(PSF) and a User-Conditioned Preference Encoder~(UCPE) to model the perception and impact of situations, respectively. We conduct experiments of applying SARE on seven backbones in various settings on two real-world datasets. Experimental results indicate that SARE improves the recommendation performances significantly compared with backbones and SOTA situation-aware baselines.

\keywords{ Situation-aware Recommendation \and Personalized Ranking \and Conditioning Neural Network.}
\end{abstract}

\section{Introduction}
\label{sec:intro}
In the era of information explosion, recommender systems~(RecSys) are essential for finding items that match users' diverse and dynamic preferences.
To better meet users' information needs, the context and features of users' interactions have been widely used for RecSys.
As our understanding of users deepens, recent research has found that users' situations, such as time, location, emotions, and activities, significantly influence their preferences~\cite{li2022towards,lin2023exploring}. 
Therefore, understanding and utilizing situations is crucial for personalized recommendations.

\begin{figure}
    \centering
    \includegraphics[width=0.9\linewidth]{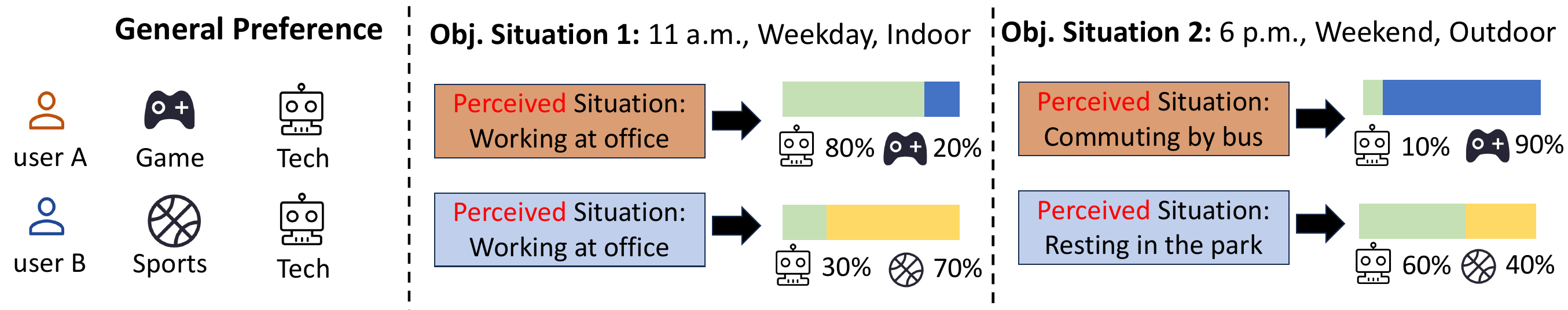}
    \caption{An example of situations as background for interactions in a news RecSys, where perceptions of situations and situations' influence on users' preference are personalized.
    }
    \label{fig:illustration}
\end{figure}

The situations we discuss include various attributes that can be obtained before users' interactions, are independent of RecSys, and reflect dynamic status of users.
For instance, environment-related and spatial-temporal features are situations, while item contents and users' in-system profiles are not.
This definition leads to two important facts of situations: 
(1) Users' preferences are sensitive to the situation: Situation is the background of interactions, influencing preferences at a higher level than inherent features of users or items, i.e., the relationship between users and items varies with situation changes.  
(2) Users' perceptions of situations are personalized, and users' preference changes with situations are also personalized~\cite{li2022towards}.
An example of two users in a news RecSys is shown in Figure~\ref{fig:illustration}.
Both users prefer technical news in general.
However, they enjoy technical news in different situations: User A prefers it at work, while User B prefers during rest.
Meanwhile, in the same objective situations, e.g., 6 p.m. outdoors at the weekend, User A usually commutes on the way, while User B is used to resting in the park, so their perceived situations are personalized, leading to different news preferences.
The example indicates that situations in RecSys influence the associations between users and items dynamically and personally, distinct from the inherent features of users and items. 

Unfortunately, both the background property and high personalization of situations are overlooked in previous RecSys.
Existing RecSys consider situations by treating them equally with other attributes or designing modules to capture the relationship between situations and user/item attributes, while they are unaware of situations' dynamic influence on user-item associations.
For instance, traditional context-aware RecSys capture situations by equally concatenating them with other attributes as input, including feature interaction models such as AFM~\cite{xiao2017attentional} and xDeepFM~\cite{lian2018xdeepfm}, and sequential models such as DIN~\cite{zhou2018deep} and DIEN~\cite{zhou2019deep}.
Some recent works focus on the connection between user/item attributes and part of situations, such as interaction intervals~\cite{wang2020make}, temporal features~\cite{cho2020meantime,ye2020time}, spatio-temporal attributes~\cite{lin2023exploring,qin2023disenpoi,lai2023multi}, and time-behavior situations~\cite{lv2023deep}.
However, they still view situations at the same level as users and items.
Therefore, they can only model the associations between situations and users/items respectively, and the situations' dynamic influences on the user-item relations have been largely neglected.
Furthermore, since situations and users are modeled together, personalized perception and influence of the situations is not considered adequately.

In this paper, we propose a new perspective that situations should serve as the precondition for recommendations rather than yet another type of feature.
Viewing situations as preconditions, we can apply distinct representations of user and item attributes given situations and better model the changes of relationships between these attributes with situation changes.
Therefore, this new perspective provides possibilities for a more comprehensive understanding of situations in RecSys. 
Meanwhile, separating situations from other attributes also provides flexibility to describe the personalized perception and influence of situations better.
Taking situations as the precondition also brings another benefit that recommendations both with and without situations can be obtained.
Thus, situations can be selectively used when they provide reliable information.

Based on this new perspective, we propose a pluggable Situation-Aware Recommender Enhancer~(SARE) module, which can flexibly incorporate situations into various existing RecSys.
In SARE, a lightweight two-tower structure is designed to find the persoanlized preferred items that matches the current situation preconditions.
SARE includes two key components:
\textit{1) User Conditional Preference Encoder}~(UCPE), a conditioning neural network to represent items conditioned on users' situation-aware personalized preference to model the personalized situation impact; 
\textit{2) Personalized Situation Fusion}~(PSF), which learns users' personalized perception of situations.
In the end, the predictions from SARE and the backbone RecSys are integrated by a combiner adaptively with the confidence of SARE and backbone outputs. 
The proposed SARE can be easily plugged in various context-aware and ID-based RecSys, enhancing the recommendation significantly with minimal additional parameters.
We conducted extensive experiments on two real-world datasets of applying SARE on five context-aware RecSys and two ID-based Recsys to confirm the effectiveness and flexibility of SARE.
Our contributions are summarized as below:
\begin{itemize}
    \item We propose a new perspective to take situations as the preconditions of interactions: Situations should be modeled separately at a higher level than users/items to capture their dynamic influences on user-item relationships.
    \item Based on the new perspective, we design SARE, a Situation-Aware Recommender Enhancer module, to model situations as preconditions considering personalized perception and impact on preferences. SARE is a pluggable module that can be applied to various RecSys to enhance their performances and provide a deeper understanding of situations.
    \item Experiments of applying SARE to five context-aware RecSys on two real-world datasets show significant improvement of SARE over backbones and SOTA methods for situation modeling. Improvements on two SARE-enhanced ID-based RecSys further indicate the flexibility of SARE.
    
\end{itemize}

\section{Related Work}
\subsection{Context-aware Recommendation}

Context-aware recommendation focuses on modeling the complex relationship between user profiles, item content, and context attributes, which has been applied to situations in some previous works.
Early works on context-aware RecSys provided various model structures for the general relations of attribute features, such as interaction terms in FM~\cite{rendle2010factorization}, a combination of FM and deep neural networks in DeepFM~\cite{guo2017deepfm} and xDeepFM~\cite{lian2018xdeepfm}, and attention-based pooling in AFM~\cite{xiao2017attentional}.
As RecSys becomes increasingly complex, users' historical interactions are also considered in context modeling.
For instance, DIN~\cite{zhou2018deep} utilized an activation unit to learn the relationships between candidate items\&context and user behavior history. DIEN~\cite{zhou2019deep} adopted an interest-evolving layer to capture user interest from history. CAN~\cite{bian2022can} enhanced sequential recommenders with Co-Action Units between target item and context and history features.
Some recent works pay special attention to the dynamic nature of historical context.
For example, dynamic graph disentangling was utilized to represent the time-evolving preferences in DisenCTR~\cite{wang2022disenctr}.
MOEF~\cite{pan2023moef} applied Fourier Transformation to time windows to describe frequent promotions changes.
Long and short-term temporal contexts were considered to model users' interest in Mojito~\cite{tran2023attention}.

Although these works have provided promising ways to model context features, they did not consider situation-related attributes specifically. 
In practice, situation features are treated equally with other attributes.
However, since situations are highly dynamic and personalized, representing them with fixed embeddings and equally trained with other attributes will lead to sub-optimal performance.
Therefore, we propose to design a module especially for situations to enhance existing context-aware recommendation models.

\subsection{Situation-aware Recommendation}

Empirical studies show that users' situations influence their preferences in RecSys~\cite{li2022towards,lin2023exploring,sun2023neighborhood}.
To understand user behaviors and improve recommendation performance, researchers have paid attention to some aspects of situations.
For example, temporal features, such as the time intervals and periodic patterns of interactions along users' history\cite{wang2020make,cho2021learning,xie2022denoising}, are widely considered in sequential recommendations.
Absolute timestamps and semantic meaning of time are also utilized, such as a mixture of multi-scale temporal embeddings~\cite{cho2020meantime}, absolute and relative time patterns~\cite{ye2020time}, and dynamic user profiles with semantic features about time~\cite{rashed2022context}.
Besides time, spatial situations are widely utilized in POI recommendation and location-based services.
Some methods focus on modeling the dynamic changes of location preference with time~\cite{wang2022modeling,lan2023spatio}, and others consider the relationship between time, locations, and users jointly~\cite{lin2023exploring,lai2023multi} or separately~\cite{qin2023disenpoi}.
One recent work~\cite{lv2023deep} aims to model the relations between situations and historical item attributes by multiple specifically designed interaction layers, where time and behavior type are used as situations.

However, these works still focused on modeling the relations between situations and user/item attributes.
As we claim in Section~\ref{sec:intro},  situations influence user-item relationship at a higher level than the inherent attributes.
We propose to consider the situations as preconditions of interactions and model their personalized perception and influence.
Moreover, all the previous research claimed to deal with one or more specific types of situations.
In SARE, all situations can be modeled without restriction, and we propose a pluggable module rather than a specific model to integrate situations with existing RecSys.

\section{Task Definition}

First, we clarify the characteristics of situations we consider in this paper: 
(1) Situations can be obtained before each interaction;
(2) Different from inherent user and item metadata, situations reflect dynamic user-side status that may change in short-term;
(3) Situations describe the current factual states, which are unrelated to the presence of a RecSys.
For instance, time, location, weather, and users' ongoning activity and emotions, are all situation features.
Whereas, user profiles, such as age and gender, users' past interactions and social networks in RecSys, and any feature about the items are out of the scope of situations.

Second, we define our situation-aware recommendation task.
Defining user preferences in the probabilistic framework aligns more closely with our view of treating situations as the precondition.
Therefore, we define the recommendation task as a ranking issue on the impression list, as the impression-based ranking task provides a clearer probability definition of negative samples than the Top-k recommendation task. To formalize, the task includes:

\textbf{Input}: a user $u$, a candidate list of $M$ items  $\{i_1,i_2,...,i_M\}$, and current situation $s$ with N attributes to describe it $\{s_1,s_2,...,s_N\}$; 
\textbf{Output}: the probability that user $u$ will interact with each item $\{p(i_m|u,s)\}_{m=1}^M$, and rank all items according to the probability $\{p(i_m|u,s)\}_{m=1}^M$.
The ground truth of the ranking list includes $M^+$ positive items and $M^-$ negative items, where $M=M^++M^-$.
Note that there is no restriction to the format or number of situation attributes $\{s_n\}_{n=1}^N$ as long as they follow our definition of situations.

\section{The Situation-Aware Recommender Enhancer}

\begin{figure*}
    \includegraphics[width=\linewidth]{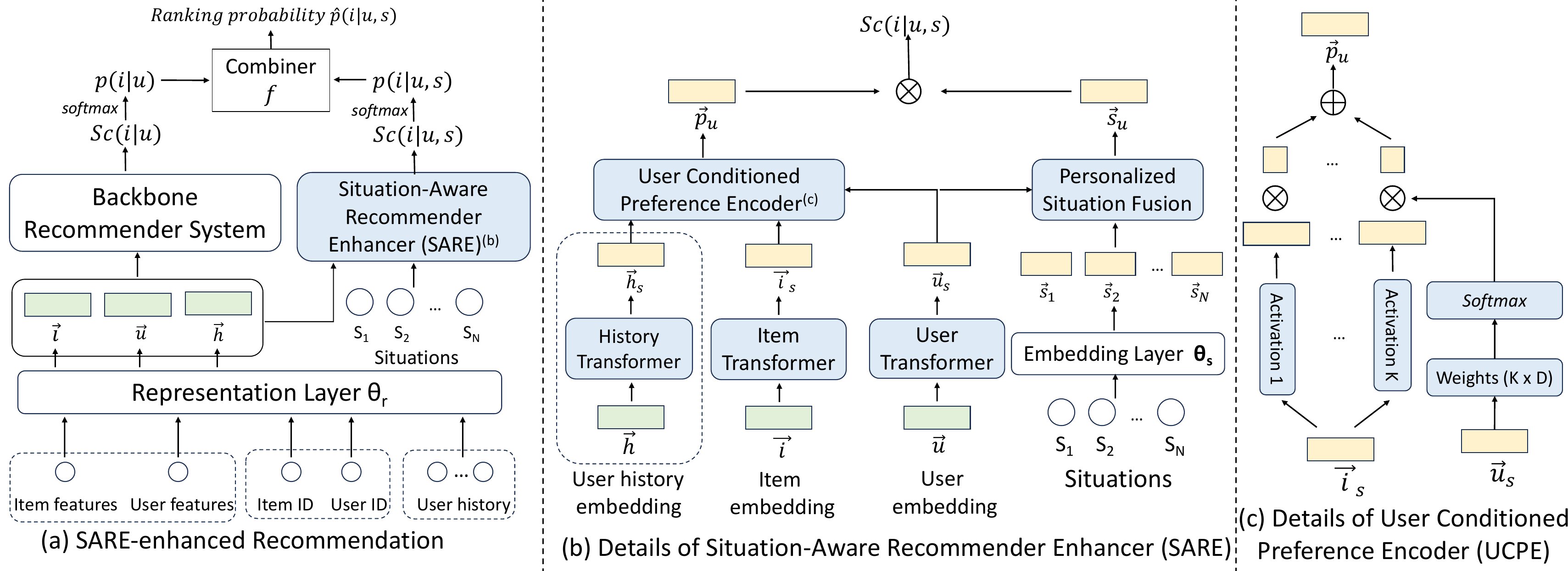}
    \caption{Overview of our proposed Situation-Oriented Optimizer~(SARE). (a) SARE can enhance backbone recommender systems with various (optional) types of input. (b) The detailed structure of SARE, where user history is an optional input. (c) The details of user-conditioned preference encoder, one of the two key components of SARE.}
    \label{fig:overall_framework}
\end{figure*}

\subsection{Overview of the Framework}
The overview of our proposed SARE Framework is shown in Figure~\ref{fig:overall_framework}(a).
SARE serves as a pluggable module to enhance various backbone RecSys with situations as interaction preconditions.
As a flexible framework, SARE enhances RecSys with one or more types of inputs, including items'/users' ID, item/user attributes, and users' interaction history, depending on the setting of backbones.
Given a user $u$, an item $i$, and optional user/item attributes and user history $h$, the inputs are embedded into vectors of dimension $D$ with a representation layer $\theta_r$ following the structure of backbone model, i.e., item $\vec i$~(including ID and attributes if any), user $\vec u$, and history $\vec h$~(optional).
The embeddings are shared between the backbone and SARE for parameter efficiency.
Then, the backbone RecSys provides a prediction score $Sc(i|u)$ without situations, and SARE will generate a prediction score $Sc(i|u,s)$ taking situations $s=\{s_1,...,s_N\}$ also as input.
Details about SARE will be illustrated in Section~\ref{subsec:method__SARE}~(Figure~\ref{fig:overall_framework}(b)). Two key components of SARE, User-Conditioned Preference Encoder~(UCPE, Figure~\ref{fig:overall_framework}(b)) and Personalized Situation Fusion~(PSF), are proposed to model the personalized influence and perception of situations, respectively, as illustrated in Section~\ref{subsec:method__condition} and Section~\ref{subsec:method__fusion}.
Finally, the output scores of the backbone $Sc(i|u)$ and SARE $Sc(i|u,s)$ are converted into probability by $softmax$ over all items in the candidate list and integrated by a combiner $f$ to form the final ranking probability $\hat p(i|u,s)$, as shown in Section~\ref{subsec:method__combiner}.

\subsection{Structure of SARE}
\label{subsec:method__SARE}
The detailed structure of SARE is shown in Figure~\ref{fig:overall_framework}(b).
It is utilized to generate users' preference for items with situations as preconditions, i.e., $p(i|u,s)$.
Firstly, since the original embeddings, i.e., item $\vec i$, user $\vec u$, and history $\vec h$~(optional), are from the non-situation space learned by the backbone recommender, we transform them into the situation-aware space for SARE,
\begin{equation}
    \vec u_s = W^{u} \vec u, \quad
    \vec i_s = W^{i} \vec i, \quad
    \vec h_s = W^{h} \vec h~(optional) 
\label{eq:transformer}
\end{equation}
Where all embedding vectors are of length $D$, and $W^u$, $W^i$, and $W^h$ are learnable matrices of size $D\times D$.
Meanwhile, $N$ situation attributes (e.g., location, time, weather) are also embedded with a linear layer $\theta_s$ into vectors $\{\vec{s}_i\}_{i=1}^N, \vec{s}_i\in \mathcal{R}^D$.

Afterward, since situations' influence on user preference and users' perception of situations are both highly personalized,
item $i_s$~(history $h_s$) and situations $\{\vec{s}_i\}_{i=1}^N$ are combined with user representation $u_s$, respectively:
UCPE, a User-Conditioned Perference Encoder, is applied for item $\vec i_s$~(and history $\vec h_s$ in sequential recommenders) and user $\vec{u}_s$ to form the user-aware preference representation $\vec{p}_u$, as shown in Section~\ref{subsec:method__condition};
PSF, a Personalized Situation Fusion, is adopted to adaptively aggregate various situation attributes $\{\vec{s}_i\}_{i=1}^N$ into a single representation $\vec s_u$ with user $\vec{u}_s$ as in Section~\ref{subsec:method__fusion}.
In the end, the predicted preference score $Sc(i|u,s)$ is generated by the inner product between personalized preference $\vec{p}_u$ and personalized situation $\vec{s}_u$, indicates the matching degree of item $i$ for user $u$ with situation $s$ as condition:

\subsection{User-Conditioned Preference Encoder~(UCPE)}
\label{subsec:method__condition}

The effect of situations on users' preferences is highly personalized.
Therefore, representations of items $\vec{i}_s$~(and history $\vec{h}_s$ for sequential recommenders) should be encoded with personalized embedding $\vec u_s$ to form personalized preference $\vec p_u$ in the situation-aware space.
Inspired by conditioning sequence modeling~\cite{ramos2021conditioning}, we propose the UCPE module as shown in Figure~\ref{fig:overall_framework}(c).

As proved in previous research~\cite{ramos2021conditioning}, a proper ensemble of various activation functions helps denoise input vectors and maintain vectors' geometry,
and learning the ensemble weights from conditioning vectors is a promising and efficient way to inject condition information~(i.e., users) into the input features~(i.e., items or history).
Taking item $\vec i_s$ and user $\vec u_s$ as an example, UCPE is applied to learn personalized weights for combining outputs of activation functions:
With an ordered group of $K$ basic activation functions, $\{A_1,A_2,...,A_K\}$~(e.g., $\{sigmoid, relu, tanh,...\}$), UCPE is designed by 
\begin{equation}
    \begin{split}
    \hat{\vec{u}} = softmax(\mathbf{W}^c\cdot \vec u_s + \vec b^c), \quad
    \vec p_u = \sum_{j=1}^K \hat u_j \cdot A_j(\vec i_s)
    \end{split}
    \label{eq:UCPE}
\end{equation}

Where $\mathbf{W}^c \in \mathcal{R}^{K\times D}$ and $\vec{b}^c\in \mathcal{R}^{K}$ are learnable parameters for personalized conditions $\hat{\vec{u}}$ on activations from the user embedding $\vec{u}_s$. $\hat{u}_j$ denotes the $j$-th demension of $\hat{\vec{u}}$.
Then personalized preference $\vec p_u$ is encoded by summation over conditions $\{\hat u_j\}$ and activated item presentations $\{A_j(\vec i_s)\}$.

For the history vector $\vec h_s$, we adopt a similar UCPE with the same activation function groups $\{A_j\}_{j=1}^K$ and different weights to generate user conditions $\hat{\vec{u}}$.
If the backbone RecSys provides both item $\vec i_s$ and history $\vec h_s$, the final preference representation $\vec p_u$ will be the sum of outputs of two UCPEs.
As we focus on proposing a lightweight situation-aware enhancer, we refrain from complex processing of user history sequences.

\subsection{Personalized Situation Fusion~(PSF)}
\label{subsec:method__fusion}

Users' perceptions of situations are also highly personalized even if the objective situation is the same.
For instance, 11 p.m. is quite late for an early riser but might still be early for a night owl.
Therefore, the representations of situations should also be personalized.
In SARE, we propose PSF~(Personalized Situation Fusion) to fuse situation attributes aware of users' personalized habits. 

Given N attributes of situations~(e.g., hour, day of week, location, weather), a cross attention module\cite{vaswani2017attention} is adopted to learn weights for situation fusion from user embedding $\vec u_s \in \mathcal{R}^{D}$ and situation embeddings $\mathbf{S}=\{\vec s_1, ... \vec s_N\} \in \mathcal{R}^{N\times D}$,
\begin{equation}
    \begin{split}
        \vec s_u = Attention(\mathbf{W}^s \vec u,\mathbf{S},\mathbf{S}) 
        =softmax\left(\frac{(\mathbf{W}^s \vec u)\cdot \mathbf{S}^T}{\sqrt{D}}\right) \cdot \mathbf{S} 
    \end{split}
    \label{eq:psf}
\end{equation}
Where $\mathbf{W}^s\in \mathcal{R}^{D\times D}$ is a learnable weight matrix.
In this way, we obtain a personalized representation $\vec s_u \in \mathcal{R}^{D}$ of the multi-attribute situations $\mathbf{S}$ considering personalized perception of user $u$ on situations.

\subsection{Probability Combiner and Model Learning}
\label{subsec:method__combiner}

For user $u$, item $i$, and situation $s$, with predicted preference scores from the backbone model $Sc(i|u)$ and SARE $Sc(i|u,s)$, we aggregate them to get a robust prediction of preference probability $\hat p(i|u,s)$ through a combiner $f$.

We combine two parts instead of directly using the SARE output $Sc(i|u,s)$ because $Sc(i|u,s)$ may not always reflect user preferences accurately due to incomplete or imprecise descriptions of users' situation $s$.
The general preference $Sc(i|u)$ helps to mitigate the impact of inaccurate modeling of situations and provides more robust recommendations.
To make the outputs on two sides comparable, we first convert the scores into probabilities $p(i|u)$ and $p(i|u,s)$ with softmax across the candidate impression list.
Then, the combiner $f$ employs a weighted harmonic mean to obtain the final probability $\hat{p}(i|u,s)$,
\begin{equation}
\label{eq:combination}
    \hat p(i|u,s) = f\left(p(i|u),p(i|u,s)\right) = \frac{c_u+c_{u,s}}{{c_u}/{p(i|u)}+{c_{u,s}}/{p(i|u,s)}}
\end{equation}
Where $c_u$ and $c_{u,s}$ are confidences of the backbone model and SARE, respectively.
Inspired by uncertainty-based learning methods~\cite{han2021trusted}, models' uncertainty is used to quantify the confidence. For the backbone model, the confidence is
\begin{equation}
    c_u = 1 - uncertainty(\{p(i_j|u)\}_{i=1}^n) = 1 - \frac{1}{\max_j \{p(i_j|u)\}_{i=1}^n+1}
\end{equation}
Meanwhile, $c_{u,s}$ is calculated in a similar way with $\{p(i_j|u,s)\}_{i=1}^n$.
Thus, we obtain a combination adaptive with the confidences of backbone and SARE. 

We adopt an end-to-end model training strategy to ensure consistency between SARE and the backbone model. 
Given a candidate list of $M^+$ positive items and $M^-$ negative items, we adopted a session-based pair-wise BPR loss~\cite{hidasi2018recurrent} for the backbone recommendation scores $\{Sc(i_m|u)\}_{m=1}^M$,
\begin{equation}
\begin{split}
    p_j^+ &= softmax_{M^+}(Sc(i_j|u)), \quad p_k^- = softmax_{M^-}(Sc(i_j|u))\\
    L_b^r&=-log\sum_{j=1}^{M^+}\sum_{k=1}^{M^-}p^+_jp^-_k\sigma\left(Sc(i_j|u)-Sc(i_k|u)\right)
\end{split}
\end{equation}
Where $p_j^+$ and $p_k^-$ are softmax probabilities over all positive and negative items, respectively, and $\sigma$ denotes Sigmoid calculation.
A similar session-level BPR loss $L_b^s$ is applied to the SARE output $\{Sc(i_m|u,s)\}_{m=1}^M$.
Meanwhile, to align the outputs of two sides, a constraint Cross-Entropy~(CE) loss $L_p$ on the final probability is applied as the optimal point of CE loss is consistent with our target on item probabilities,
\begin{equation}
    L_p = \frac{1}{M^+}\sum_{j=1}^{M^+}log\left(\hat{p}(i_j|u,s)\right) + \frac{1}{M^-}\sum_{k=1}^{M^+}log\left(1-\hat{p}(i_k|u,s)\right)
\end{equation}
The final loss is weighted sum of three parts with hyper-parameters $\lambda_s$ and $\lambda_p$,
\begin{equation}
    L = L_b^r + \lambda_s L_b^s + \lambda_p L_p
\end{equation}

Since the varying settings and complexity of the backbone RecSys leads to different optimal learning rates, we define two learning rates for parameters of SARE and the backbone RecSys, denoted as $lr_s$ and $lr_r$, respectively.

\subsection{Discussions on Parameter Efficiency}

As a pluggable module, SARE enhances the backbone RecSys with negligible additional parameters compared with the embedding parameters.
Let the embedding dimension be $D$, parameters for transformers add up to at most $3D^2$ as in Eq\ref{eq:transformer}, the UCPE module includes at most $2K(D+1)$ parameters in Eq\ref{eq:UCPE}, and the PSF contains attention matrix with $D^2$ parameters as in Eq\ref{eq:psf}.
In all, the additional parameters are at most $4D^2+2KD+2K$, which is much smaller than
the embedding matrix for users and items of backbone RecSys with $D(|\mathcal{U}|+|\mathcal{I}|)$ parameters, since generally $(D+K)<<|\mathcal{U}|+|\mathcal{I}|$.

\section{Experiments}

\subsection{Experimental Settings}

\subsubsection{Datasets}
\begin{table}[]
\caption{Dataset statistics. 
\# indicates \textit{the number of}, Imp. is short for \textit{impression}, and a. is short for \textit{attributes}.
}
\label{tab:dataset}
\begin{tabular}{c|ccccc}
\hline
\textbf{Dataset} & \textbf{\#User} & \textbf{\#Item} & \textbf{\#Click} & \textbf{\#Imp. list} & \textbf{Click/Imp.} \\
\hline
\textbf{MIND} & 40,000 & 25,545 & 332,756 & 309,134 & 1.076 \\
\textbf{KuaiRand-1k} & 1,000 & 487,974 & 2,763,334 & 201,889 & 12.430 \\
\hline
\textbf{Dataset} & \textbf{Imp. length} & \textbf{\#User a.} & \textbf{\#Item a.} & \textbf{\#Situation a.} &  \\
\hline
\textbf{MIND} & 46.922 & 1 & 3 & 4 &  \\
\textbf{KuaiRand-1k} & 29.639 & 25 & 7 & 4 & \\
\hline
\end{tabular}
\end{table}

Since the situation-aware recommendation task is defined as ranking on the impression item lists, we choose two large-scale public recommendation datasets with impression records: a news dataset, MIND~\cite{wu2020mind}, and a short video dataset, KuaiRand-1K~\cite{gao2022kuairand}.
As users are more sensitive to situations when consuming information than purchasing products~\cite{li2022towards}, both datasets are information streaming datasets, while they focus on distinct scenarios.
For the ranking list, MIND includes impression logs in the original records, and in KuaiRand-1K, we split each user's records into impression lists with an interval of 30 minutes.
For both datasets, we filter impression lists including users and items with less than three positive interactions.
Moreover, we randomly select 40,000 users from all filtered users in MIND to get a dataset with a proper size.
We adopt all user and item attributes in the dataset for context-aware backbones and baselines.
The statistics of the two datasets are shown in Table~\ref{tab:dataset}.

For situation attributes, we obtain four semantic time-related attributes following Lv et al.~\cite{lv2023deep}: hour of day, day of week, period of day, and weekend or not.
Note that time-related situations are utilized since situation information is limited in existing datasets and temporal pattern has great influence on user preferences~\cite{lv2023deep},
but SARE can be applied to any kind of situations without constraint.
Moreover, to further illustrate the flexibility of SARE for situations, we retain dependent situation attributes~(i.e., hour and period, day of week and weekend or not) for the model to learn their relationship automatically.

\subsubsection{Evaluation Protocols}
We randomly split all impression item lists into training, validation, and test sets in an 8:1:1 ratio.
Random rather than timeline split is adopted to ensure that situations follow the same distribution in training and test stage for a fair comparison.
The ground truth is to rank all clicked~(i.e., positive) items higher than all non-click items in an impression list.
HR@3, MAP@3, and NDCG@3\cite{jarvelin2002cumulated} are utilized as evaluation metrics.
We repeat each experiment 5 times with different random seeds and report the average values.

\subsubsection{Context-aware Recommendation}

Firstly, we evaluate SARE with context-aware RecSys as backbones. We select backbones to include well-known models and diverse model structures, including both non-sequential and sequential ones:

1) \textbf{FM}~\cite{rendle2010factorization} is one of the most classic context-aware RecSys to combine item, user, and context attributes in first and second orders. 
2) \textbf{AFM}~\cite{xiao2017attentional} uses attention architectures to learn the importance of different feature combinations.
3) \textbf{xDeepFM}~\cite{lian2018xdeepfm} aims to explicitly generate feature interactions at the vector level with a proposed Compressed Interaction Network~(CIN).
4) \textbf{DIN}~\cite{zhou2018deep} is a sequential context-aware RecSys that proposes local activation units to assign weights for users' historical interactions dynamically.
5) \textbf{CAN}~\cite{bian2022can} is a SOTA context-aware RecSys module, which disentangles the historical representation and feature interactions with a Co-Action Network, where we adopt DIN as the backbone for CAN, and item id is used as the induction feature.
For each model, a SARE-enhanced version is compared with the original backbone.
For a fair comparison, situation attributes are concatenated with user\&item features for backbones without SARE, and removed from the backbone sides when adding SARE.

Moreover, we compare SARE-enhanced models with some SOTA situation-aware recommenders.
Since few models deal with general situation attributes, we consider context-aware RecSys that include temporal features in two ways:

(1) Model the dynamic changes of user preference or context features:
\textbf{DIF-SR}~\cite{xie2022decoupled} aimed to fuse dynamic context for sequential RecSys with a decoupled attention mechanism and a context prediction module; 
\textbf{CARCA}~\cite{rashed2022context} utilized multi-level attention components to capture the dynamic influence of users' historical attributes for the current interactions.

(2) Model time as semantic features explicitly: 
\textbf{MEANTIME}~\cite{cho2020meantime} utilized absolute and relative embeddings of time sequence for sequential recommendation;
\textbf{MOJITO}~\cite{tran2023attention} proposed a time-aware sequential RecSys with temporal context embeddings using a Mercer kernel function.

For all the above methods, we utilize all attributes of user, item, and context in SARE as input for a fair comparison.
Note that although DSAIN~\cite{lv2023deep} is claimed as a situation-aware RecSys, it focuses on multi-behavior scenario.
We attempted to reproduce i on our single-behavior datasets, but did not achieve desired performances.
Therefore, it is not included to avoid unfair comparisons.

\subsubsection{ID-based Recommendation}
To further illustrate the flexibility of our proposed SARE framework, we apply SARE to ID-based RecSys, where context features are initially not considered.
This suits when there exists a RecSys relying on ID information, and we are attempting to incorporate the influence of situations.
Therefore, we consider two settings: \textbf{fix+SARE}, an ID-based backbone is pre-trained without situation information and fixed, and SARE is then trained with situation-aware data on the fixed backbone; \textbf{train+SARE}, SARE and backbones are trained together.
We select two powerful well-known ID-based RecSys as backbones to indicate the universal utility of SARE:
\textbf{LightGCN}~\cite{he2020lightgcn} is a graph-based non-sequential RecSys that applies a light graph convolution module to model user-item interactions;
\textbf{SASRec}~\cite{kang2018self} is a sequential RecSys utilizing self-attention blocks to predict the next item. 

\subsubsection{Implementation Details}
We implement all backbones, baselines, and our proposed SARE with Pytorch.
We adopt Adam as the optimizer for all models and tune learning rates for each method.
We fix embedding sizes $D$ for all models as 64 for a fair comparison except CAN, which requires a larger embedding size for the induction feature, i.e., item id.
For our proposed SARE module, 11 activation functions are defined for UCPE following previous work~\cite{ramos2021conditioning}:
We tune SARE-specific learning rate, weight decay, and loss weights $\lambda_s,\lambda_p$ for SARE with grid search.
Source codes and detailed hyper-parameters are released at \href{https://github.com/JiayuLi-997/SARE\_DASFFA2024}{https://github.com/JiayuLi-997/SARE\_DASFFA2024}.

\subsection{Enhancement with SARE on Context-aware Recommendation}
\begin{table}[]
\caption{Performances on five SARE-enhanced context-aware RecSys. Each SARE-enhanced model is compared with the corresponding backbones, where boldface shows the best in the group, and */** denotes significant differences with $p<0.05/0.01$ by paired t-test. For fair comparison, situations are used as context features for all backbones, and removed from the backbone sides when adding SARE.}
\resizebox{\columnwidth}{!}{
\begin{tabular}{cc|ccc|ccc}
\hline
\multirow{2}{*}{\textbf{Models}} & \multirow{2}{*}{\textbf{Situation}} & \multicolumn{3}{c|}{\textbf{MIND}} & \multicolumn{3}{c}{\textbf{KuaiRand}} \\
& & \textbf{HR@3} & \textbf{MAP@3} & \textbf{NDCG@3} & \textbf{HR@3} & \textbf{MAP@3} & \textbf{NDCG@3} \\
 \hline
\textbf{FM}~\cite{rendle2010factorization} & \textbf{Concat}& 0.4774 & 0.3011 & 0.3407 & 0.8717 & 0.5507 & 0.6221 \\
 & \textbf{SARE} & \textbf{0.4819}* & \textbf{0.3028}* & \textbf{0.3444}** & \textbf{0.8771}** & \textbf{0.5605}** & \textbf{0.6308}** \\
\hline
\textbf{AFM}~\cite{xiao2017attentional} & \textbf{Concat} & 0.4780 & 0.2997 & 0.3398 & {0.8758} & {0.5551} & {0.6266} \\
 & \textbf{SARE} & \textbf{0.4797} & \textbf{0.3015}*  & \textbf{0.3417}* & \textbf{0.8789}* & \textbf{0.5602}* & \textbf{0.6301}* \\ \hline
\textbf{xDeepFM}~\cite{lian2018xdeepfm} & \textbf{Concat} & {0.4825} & {0.3055} & {0.3453} & {0.8733} & {0.5512} & {0.6226} \\
& \textbf{SARE} & \textbf{0.4857}*  & \textbf{0.3114}** & \textbf{0.3504}* & \textbf{0.8832}**  & \textbf{0.5705}**  & \textbf{0.6405}** \\ \hline
\textbf{DIN}~\cite{zhou2018deep} & \textbf{Concat} & \textbf{0.4834} & 0.3038 & 0.3441 & 0.8700 & 0.5454 & 0.6174 \\
 & \textbf{SARE} & 0.4821 & \textbf{0.3057}* & \textbf{0.3470} & \textbf{0.8761}* & \textbf{0.5489} & \textbf{0.6203} \\ \hline
\textbf{CAN}~\cite{bian2022can} & \textbf{Concat} & 0.4850 & 0.3055 & 0.3459 & 0.8749 & 0.5496 & 0.6218 \\
& \textbf{SARE}  & \textbf{0.4881}  & \textbf{0.3089}* & \textbf{0.3493}* & \textbf{0.8762} & \textbf{0.5610}** & \textbf{0.6310}** \\
\hline
\end{tabular}
}
\label{tab:all_results}
\end{table}

The results of comparisons between context-aware RecSys and SARE-enhanced ones are shown in Table~\ref{tab:all_results}.
With the same information as input, SARE leads to significantly better performance than concatenating situation features on most backbones for both datasets, indicating the efficacy of SARE for dealing with situations.
Specifically, adding SARE to the simple FM backbone outperforms all backbones for KuaiRand, showing that our new perspective of viewing situations as preconditions helps more than complex modeling of the relationship between situations and other attributes at the same level.
xDeepFM+SARE performs the best on most metrics for both datasets, even surpassing sequential methods.
It may be because feature associations within the impression play a crucial role in the setting of impression list ranking.
Based on xDeepFM, leveraging situations as additional conditions allows for a more detailed consideration of how feature interactions change under the influence of different situations, leading to the best performance.
Notably, on KuaiRand, non-sequential methods perform well, and the incorporation of SARE consistently provides substantial improvements. 
It indicates that in this scenario, features' semantic information may be more crucial than user history, emphasizing the importance of SARE for capturing the personalized semantic meaning and influence of situations.
We also find that SARE leads to more performance improvements on non-sequential models than sequential RecSys, DIN and CAN. It may be because SARE is designed mainly for user-item interactions, refraining from complex processing of user history. Design of an enhanced situation-aware model that better integrates user history is anticipated to yield more improvements on sequential Recsys.

\begin{figure}
    \centering
    \includegraphics[width=0.9\linewidth]{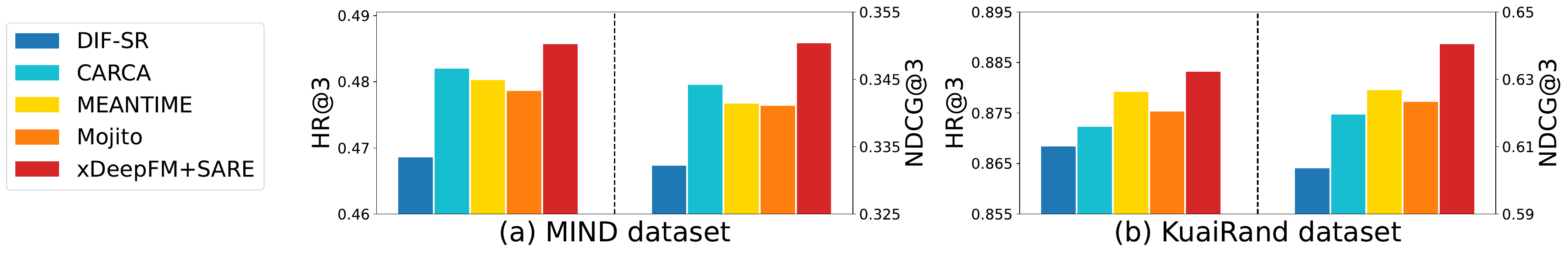}
    \caption{Comparisons between xDeepFM+SARE and SOTA situation~(time)-aware models. xDeepFM+SARE outperforms all baselines significantly.}
    \label{fig:baseline}
\end{figure}

Results of the best SARE-enhanced model and SOTA baselines are shown in Figure~\ref{fig:baseline}.
xDeepFM+SARE shows a significant performance improvement compared to all baseline models.
Comparing different baselines, we observe that DIF-SR, which directly predicts situations, struggles to learn effectively since numerous situations exist in our experiments.
On the MIND dataset, the importance of sequential information results in the good performance of CARCA. On KuaiRand, two semantic-based baselines, MEANTIME and Mojito, perform better because semantic time features are crucial.
Finally, by thoroughly considering the personalized semantic meanings and influences of situations without restrictions, our proposed SARE achieves the best performance.

\subsection{Ablation Study of SARE on Context-aware RecSys}

\begin{figure}
    \centering
    \includegraphics[width=0.9\linewidth]{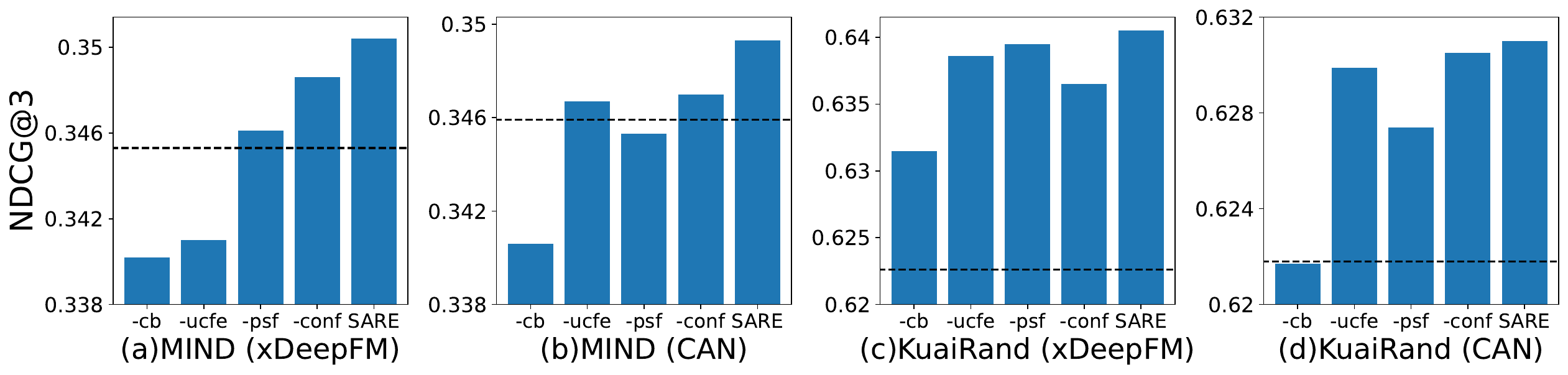}
    \caption{Ablation study with SARE and its variant, without: combiner(-cb., using SARE output only), ucfe, psf, confidence-aware weights(-conf., harmonic mean without weights). The dotted lines denote backbone performances. }
    \label{fig:ablation}
\end{figure}

The ablation study of SARE with xDeepFM~(non-sequential) and CAN~(sequential) as backbones on both datasets is shown in Figure~\ref{fig:ablation}.
Variants without UCFE or PSF~(i.e., -ucfe and -psf) are conducted by replacing two personalized components with global weights learnt the same for all users.
We also consider a variant that exclude backbone outputs~(i.e., -cb), which use $p(i|u,s)$ rather than $\hat{p}(i|u,s)$ in Eq.~\ref{eq:combination} as the final output of model, and
a variant that take harmonic means of $p(i|u)$ and $p(i|u,s)$ without confidences in Eq.~\ref{eq:combination}, \textit{-conf}.
The results show that both components contribute to the performances, especially for the MIND dataset.
Therefore, personalization of situation perceptions and influences are important for situation modeling.
Moreover, the integration of SARE and backbone outputs, as well as the integration strategy, are also crucial. 
Notably, excluding backbone outputs~(i.e., -cb) leads to the most performance decay, resulting from a lack of direct supervision signal for the backbone and noises in the situation modeling. 
Both SARE and backbone optimization are sub-optimal in the \textit{-cb} setting.
Combinations aware of the confidence of SARE and backbone outputs, i.e., SARE v.s. -conf, further enhances the models' performance.

\subsection{Enhancement with SARE on ID-based Recommendation}

\begin{table}[h]
\caption{Performances of SARE-enhanced ID-based recommendation with/without pre-trained backbones and ID-based backbone models. All SARE methods are compared with the corresponding backbones. Other notations are the same as Table~\ref{tab:all_results}.}
\resizebox{\columnwidth}{!}{
\label{tab:ID_base}
\begin{tabular}{c|c|ccc|ccc}
\hline
\multirow{2}{*}{\textbf{Backbone}} & \multirow{2}{*}{\textbf{Setting}} & \multicolumn{3}{c|}{\textbf{MIND}} & \multicolumn{3}{c}{\textbf{KuaiRand}} \\
 &  & \textbf{HR@3} & \textbf{MAP@3} & \textbf{NDCG@3} & \textbf{HR@3} & \textbf{MAP@3} & \textbf{NDCG@3} \\
 \hline
\multirow{3}{*}{\textbf{\begin{tabular}[c]{@{}c@{}}LightGCN\\ {\cite{he2020lightgcn}}\end{tabular}}}
& {w/o situation} & 0.4471 & 0.2735 &0.3125  & 0.8589  & 0.5324  & 0.6043  \\
 & {fix + SARE} & 0.4514** & 0.2779** & 0.3167** & \textbf{0.8672}**  & {0.5411}** & {0.6130}**  \\
 & {train + SARE}& \textbf{0.4599}** & \textbf{0.2796}** & \textbf{0.3184}** & {0.8661}**  & \textbf{0.5419}** & \textbf{0.6137}** \\
 \hline
 \multirow{3}{*}{\textbf{\begin{tabular}[c]{@{}c@{}}SASRec\\ {\cite{kang2018self}}\end{tabular}}}
& {w/o situation} & 0.4670 & 0.2905 & 0.3301 & 0.8594  & 0.5279 & 0.6023 \\
 & {fix + SARE} & {0.4688} & {0.2934}* & {0.3328} & 0.8611  & 0.5299 & 0.6026 \\
 & {train + SARE} & \textbf{0.4757}**  & \textbf{0.2986}** & \textbf{0.3385}** & \textbf{0.8643}* & \textbf{0.5328}** & \textbf{0.6052}*  \\
 \hline
\end{tabular}
}
\end{table}

The results of SARE-enhanced ID-based RecSys are shown in Table~\ref{tab:ID_base}.
SARE trained on fixed backbones outperforms the original backbones, and training backbone and SARE together yields even better results, suggesting that SARE can learn situation-aware information based on both existing ID embeddings and adjustable embeddings. 
Therefore, as a pluggable module, SARE is able to enhance models initially without any context features, showing its substantial flexibility.
On the other hand, ID-based methods underperform context-aware approaches, which shows that modeling situations with SARE benefits from additional semantic information for items and users since situations are considered as preconditions for modeling relations of user\&item features in SARE.

\section{Conclusion and Future Work}
In this paper, we propose a new perspective of viewing situations as the preconditions of interactions in the recommender systems, which helps understand the dynamic and personalized influence of situations on user-item relationships.
Based on this, a novel Situation-Aware Recommender Enhancer is designed to model situations as preconditions.
SARE is a flexible and pluggable module to integrate situations into various existing RecSys.
As users' perceptions of situations and the impact of situations on preference are both highly personalized, we design a personalized situation fusion and a user-conditioned preference encoder to model the perception and influence of situations, respectively.
Then, the recommendations from SARE and backbone model are integrated by a combiner considering the model confidence.
Extensive experiments with five context-aware RecSys and two ID-based RecSys as backbones show significant improvements with SARE to integrate situations on two real-world recommendation datasets.

In the future, we will combine SARE with more SOTA context-aware RecSys and consider more aspects of situations in experiments to further verify its effectiveness.
Moreover, as SARE focuses on dealing with situations separately as preconditions, utilization of it in sequential RecSys is straightforward.
Better adapting SARE for sequential Recsys will be explored in the next step.

\subsubsection{\ackname} This work is supported by the Natural Science Foundation of China (Grant No. U21B2026, 62372260), Quan Cheng Laboratory (Grant No. QCLZD202301), and the fellowship of China Postdoctoral Science Foundation (No.2022TQ0178).

%
%
%
\bibliographystyle{splncs04}
\bibliography{reference}

\begin{thebibliography}{10}
\providecommand{\url}[1]{\texttt{#1}}
\providecommand{\urlprefix}{URL }
\providecommand{\doi}[1]{https://doi.org/#1}

\bibitem{bian2022can}
Bian, W., Wu, K., Ren, L., Pi, Q., Zhang, Y., Xiao, C., Sheng, X.R., Zhu, Y.N., Chan, Z., Mou, N., Luo, X., Xiang, S., Zhou, G., Zhu, X., Deng, H.: Can: Feature co-action network for click-through rate prediction. In: Proceedings of the 15th ACM WSDM. p. 57–65. WSDM '22, New York, NY, USA (2022)

\bibitem{cho2021learning}
Cho, J., Hyun, D., Kang, S., Yu, H.: Learning heterogeneous temporal patterns of user preference for timely recommendation. In: Proceedings of the Web Conference 2021. pp. 1274--1283 (2021)

\bibitem{cho2020meantime}
Cho, S.M., Park, E., Yoo, S.: Meantime: Mixture of attention mechanisms with multi-temporal embeddings for sequential recommendation. In: Proceedings of the 14th ACM Conference on Recommender Systems. pp. 515--520 (2020)

\bibitem{gao2022kuairand}
Gao, C., Li, S., Zhang, Y., Chen, J., Li, B., Lei, W., Jiang, P., He, X.: Kuairand: An unbiased sequential recommendation dataset with randomly exposed videos. In: Proceedings of the 31st ACM CIKM. p. 3953–3957 (2022)

\bibitem{guo2017deepfm}
Guo, H., Tang, R., Ye, Y., Li, Z., He, X.: Deepfm: a factorization-machine based neural network for ctr prediction. arXiv preprint arXiv:1703.04247  (2017)

\bibitem{han2021trusted}
Han, Z., Zhang, C., Fu, H., Zhou, J.T.: Trusted multi-view classification. arXiv preprint arXiv:2102.02051  (2021)

\bibitem{he2020lightgcn}
He, X., Deng, K., Wang, X., Li, Y., Zhang, Y., Wang, M.: Lightgcn: Simplifying and powering graph convolution network for recommendation. In: Proceedings of the 43rd International ACM SIGIR. pp. 639--648 (2020)

\bibitem{hidasi2018recurrent}
Hidasi, B., Karatzoglou, A.: Recurrent neural networks with top-k gains for session-based recommendations. In: Proceedings of the 27th ACM international conference on information and knowledge management. pp. 843--852 (2018)

\bibitem{jarvelin2002cumulated}
J{\"a}rvelin, K., Kek{\"a}l{\"a}inen, J.: Cumulated gain-based evaluation of ir techniques. ACM Transactions on Information Systems (TOIS)  \textbf{20}(4),  422--446 (2002)

\bibitem{kang2018self}
Kang, W.C., McAuley, J.: Self-attentive sequential recommendation. In: 2018 IEEE international conference on data mining (ICDM). pp. 197--206. IEEE (2018)

\bibitem{lai2023multi}
Lai, Y., Su, Y., Wei, L., Chen, G., Wang, T., Zha, D.: Multi-view spatial-temporal enhanced hypergraph network for next poi recommendation. In: International Conference on Database Systems for Advanced Applications. pp. 237--252 (2023)

\bibitem{lan2023spatio}
Lan, P., Zhang, Y., Xiang, H., Wang, Y., Zhou, W.: Spatio-temporal position-extended and gated-deep network for next poi recommendation. In: International Conference on Database Systems for Advanced Applications. pp. 505--520 (2023)

\bibitem{li2022towards}
Li, J., He, Z., Cui, Y., Wang, C., Chen, C., Yu, C., Zhang, M., Liu, Y., Ma, S.: Towards ubiquitous personalized music recommendation with smart bracelets. Proceedings of the ACM IMWUT  \textbf{6}(3),  1--34 (2022)

\bibitem{lian2018xdeepfm}
Lian, J., Zhou, X., Zhang, F., Chen, Z., Xie, X., Sun, G.: xdeepfm: Combining explicit and implicit feature interactions for recommender systems. In: Proceedings of the 24th ACM SIGKDD. pp. 1754--1763 (2018)

\bibitem{lin2023exploring}
Lin, S., Pei, J., Zhou, T., He, H., Jia, J., Hu, N.: Exploring the spatiotemporal features of online food recommendation service. In: Proceedings of the 46th International ACM SIGIR. pp. 3354--3358 (2023)

\bibitem{lv2023deep}
Lv, Y., Wang, S., Jin, B., Yu, Y., Zhang, Y., Dong, J., Wang, Y., Wang, X., Wang, D.: Deep situation-aware interaction network for click-through rate prediction. In: Proceedings of the 17th ACM Conference on Recommender Systems (2023)

\bibitem{pan2023moef}
Pan, X., Shen, Y., Zhang, J., He, X., Huang, Y., Wen, H., Mao, C., Cao, B.: Moef: Modeling occasion evolution in frequency domain for promotion-aware click-through rate prediction. In: International Conference on Database Systems for Advanced Applications. pp. 330--340 (2023)

\bibitem{qin2023disenpoi}
Qin, Y., Wang, Y., Sun, F., Ju, W., Hou, X., Wang, Z., Cheng, J., Lei, J., Zhang, M.: Disenpoi: Disentangling sequential and geographical influence for point-of-interest recommendation. In: Proceedings of the 16th ACM WSDM (2023)

\bibitem{ramos2021conditioning}
Ramos, A.G.C.P., Mehrotra, A., Lane, N.D., Bhattacharya, S.: Conditioning sequence-to-sequence networks with learned activations. In: International Conference on Learning Representations (2021)

\bibitem{rashed2022context}
Rashed, A., Elsayed, S., Schmidt-Thieme, L.: Context and attribute-aware sequential recommendation via cross-attention. In: Proceedings of the 16th ACM Conference on Recommender Systems. pp. 71--80 (2022)

\bibitem{rendle2010factorization}
Rendle, S.: Factorization machines. In: 2010 IEEE International conference on data mining. pp. 995--1000. IEEE (2010)

\bibitem{sun2023neighborhood}
Sun, P., Wu, L., Zhang, K., Chen, X., Wang, M.: Neighborhood-enhanced supervised contrastive learning for collaborative filtering. IEEE Transactions on Knowledge and Data Engineering  (2023)

\bibitem{tran2023attention}
Tran, V.A., Salha-Galvan, G., Sguerra, B., Hennequin, R.: Attention mixtures for time-aware sequential recommendation. In: Proceedings of the 46th International ACM SIGIR. pp. 1821--1826 (2023)

\bibitem{vaswani2017attention}
Vaswani, A., Shazeer, N., Parmar, N., Uszkoreit, J., Jones, L., Gomez, A.N., Kaiser, {\L}., Polosukhin, I.: Attention is all you need. Advances in neural information processing systems  \textbf{30} (2017)

\bibitem{wang2020make}
Wang, C., Zhang, M., Ma, W., Liu, Y., Ma, S.: Make it a chorus: knowledge-and time-aware item modeling for sequential recommendation. In: Proceedings of the 43rd International ACM SIGIR. pp. 109--118 (2020)

\bibitem{wang2022modeling}
Wang, X., Sun, G., Fang, X., Yang, J., Wang, S.: Modeling spatio-temporal neighbourhood for personalized poi recommendation. In: Proceedings of IJCAI (2022)

\bibitem{wang2022disenctr}
Wang, Y., Qin, Y., Sun, F., Zhang, B., Hou, X., Hu, K., Cheng, J., Lei, J., Zhang, M.: Disenctr: Dynamic graph-based disentangled representation for click-through rate prediction. In: Proceedings of the 45th International ACM SIGIR (2022)

\bibitem{wu2020mind}
Wu, F., Qiao, Y., Chen, J.H., Wu, C., Qi, T., Lian, J., Liu, D., Xie, X., Gao, J., Wu, W., et~al.: Mind: A large-scale dataset for news recommendation. In: Proceedings of the 58th Annual Meeting of ACL. pp. 3597--3606 (2020)

\bibitem{xiao2017attentional}
Xiao, J., Ye, H., He, X., Zhang, H., Wu, F., Chua, T.S.: Attentional factorization machines: Learning the weight of feature interactions via attention networks. arXiv preprint arXiv:1708.04617  (2017)

\bibitem{xie2022denoising}
Xie, S., Li, Q., Xu, W., Shen, K., Chen, S., Zhong, W.: Denoising time cycle modeling for recommendation. In: Proceedings of the 45th International ACM SIGIR. pp. 1950--1955 (2022)

\bibitem{xie2022decoupled}
Xie, Y., Zhou, P., Kim, S.: Decoupled side information fusion for sequential recommendation. In: Proceedings of the 45th International ACM SIGIR (2022)

\bibitem{ye2020time}
Ye, W., Wang, S., Chen, X., Wang, X., Qin, Z., Yin, D.: Time matters: Sequential recommendation with complex temporal information. In: Proceedings of the 43rd international ACM SIGIR. pp. 1459--1468 (2020)

\bibitem{zhou2019deep}
Zhou, G., Mou, N., Fan, Y., Pi, Q., Bian, W., Zhou, C., Zhu, X., Gai, K.: Deep interest evolution network for click-through rate prediction. In: Proceedings of the AAAI conference on artificial intelligence. vol.~33, pp. 5941--5948 (2019)

\bibitem{zhou2018deep}
Zhou, G., Zhu, X., Song, C., Fan, Y., Zhu, H., Ma, X., Yan, Y., Jin, J., Li, H., Gai, K.: Deep interest network for click-through rate prediction. In: Proceedings of the 24th ACM SIGKDD. pp. 1059--1068 (2018)

\end{thebibliography}
\end{document}